\documentclass[aps,prd,10pt,twocolumn,showpacs,preprintnumbers,amsmath,amssymb,nofootinbib,floatfix]{revtex4-1}

\usepackage{natbib} 
\usepackage{color}
\usepackage{amsmath}
\usepackage{epsfig}
\usepackage{array}
\usepackage[hyperfootnotes=true]{hyperref} 
\usepackage{afterpage}

\usepackage{natbib}
\usepackage{graphicx}

\begin{document}

\title{A novel approach for using programming exercises in electromagnetism coursework}

\author{Orban, C.$^*$}
\affiliation{Department of Physics, The Ohio State University, Columbus, OH, 43210}
\author{Porter, C. D}
\affiliation{Department of Physics, The Ohio State University, Columbus, OH, 43210}
\author{Brecht, N. K.}
\affiliation{Department of Physics, The Ohio State University, Columbus, OH, 43210}
\author{Teeling-Smith, R. M.}
\affiliation{Marion Technical College, Marion, OH, 43302}
\affiliation{University of Mt. Union, Alliance, OH, 44601}
\author{Harper, K. A.}
\affiliation{Department of Engineering Education, The Ohio State University, Columbus, OH, 43210}

\email{orban@physics.osu.edu}

\date{\today}

\begin{abstract}
While there exists a significant number of web interactives for introductory physics, students are almost never shown the computer code that generates these interactives even when the physics parts of these programs are relatively simple. Building off of a set of carefully-designed classical mechanics programming exercises that were constructed with this goal in mind, we present a series of electromagnetism programming exercises in a browser-based framework called p5.js. Importantly, this framework can be used to highlight the physics aspects of an interactive simulation code while obscuring other details. This approach allows absolute beginner programmers to gain experience in modifying and running the program without becoming overwhelmed. We plan to probe the impact on student conceptual learning using the Brief Electricity and Magnetism Assessment and other questions. We invite collaborators and teachers to adopt this framework in their high school or early undergraduate classes. All exercises are available at \href{http://compadre.org/PICUP}{compadre.org/PICUP}
\end{abstract}

\maketitle

\section{Introduction}

As discussed in \cite{Orban_etal2017,Sherin2001,Landau_etal2011} and references therein, there is a compelling need to incorporate computer programming content into introductory physics, and the environment for doing so is as supportive as it has ever been. Changes in federal legislation prompted by the group code.org and statements regarding the importance of ``computational thinking" in the Next Generation Science Standards (NGSS, \cite{NGSS}) imply that teachers can freely use computer programming activities in STEM courses so long as the content of these activities pertains to concepts and material that are ordinarily covered in that course. With this in mind, \citet{Orban_etal2017} presented a set of computer programming activities for classical mechanics that were designed for students with no prior programming experience. These activities were designed on the premise that students need an approach that is (1) simple, involving 75 lines of code or fewer with plenty of comments, (2) easy to use with browser-based coding tools (3) interactive, with a high frame rate to give a video-game like feel, (4) step-by-step, with the ability to interact with intermediate stages of the "correct" program and (5) thoughtfully integrated into the physics curriculum, for example, by illustrating velocity and acceleration vectors throughout. As a continuation of \cite{Orban_etal2017}, we present a set of programming activities for introductory electromagnetism courses. Similar to \cite{Orban_etal2017}, these activities have been classroom tested at OSU's regional campus in Marion, Ohio.

\section{Overview of Programming Activities for Electromagnetism}

In the second semester of introductory physics at Ohio State University at the regional campus in Marion, we include six required programming activities. The official description of this course is calculus-based physics II, but at OSU students only need to have finished the first semester of calculus (which does not include integrals) to enroll in the course. As a result the calculus content in the course is rather limited, and the programming exercises reflect this limitation. 

The list of exercises is as follows: (1) Particle Accelerator!, (2) Particle Accelerator (with potential!), (3) Repulsion between two point charges, (4) RC circuits, (5) Magnetic Force, and (6) Wave Interference. Each activity is designed to take about an hour to complete. All of these exercises and solutions are available at \href{http://compadre.org/PICUP}{compadre.org/PICUP}.   Because of the algebra-based physics emphasis of these exercises we regard these exercises to be appropriate for high school and early college introductory physics, and the exercises work well on chromebooks or any computer with a keyboard.  

\begin{figure*}
\includegraphics[width=6in]{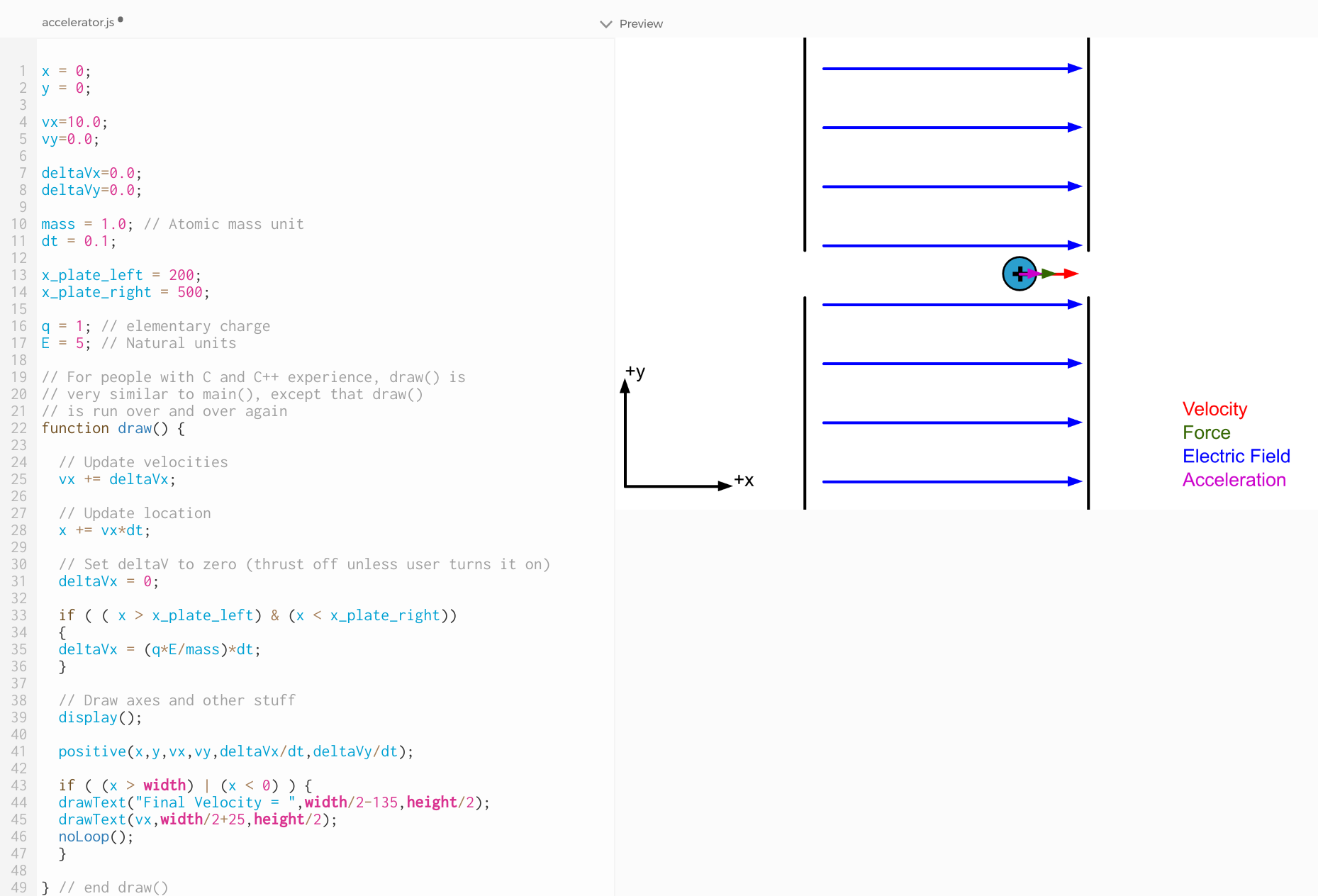}
\caption{A screenshot of the Particle Accelerator programming activity from the in-browser interface. The visualization on the right shows the particle being accelerated from between two charged plates. The particle enters from the left traveling slowly and exits to the right.} \label{fig:accelerator}
\end{figure*}

\section{Exercises 1 \& 2: Particle Accelerator}

The first two exercises use exactly the same code and the situation involves accelerating an initially slow moving charged particle using two charged plates (Fig.~\ref{fig:accelerator}). Fig.~\ref{fig:accelerator} is itself a screenshot from an in-browser coding environment. Importantly, there are fewer than 50 lines of code and each section is well commented. The code is structured in much the same way as the classical mechanics exercises, as discussed in \cite{Orban_etal2017}, with global variable initializations at the top, a \texttt{draw()} function that runs many times per second, and within the draw function the velocity and position update is first, followed by code that determines the acceleration and change in velocity for the next step\footnote{As in \cite{Orban_etal2017}, and as a careful look at Fig.~\ref{fig:accelerator} reveals, the velocity and position updates follow an Euler-Cromer method \cite{Cromer1981}.}. All other code is hidden in the subroutine \texttt{functions()} that the student does not need to see to complete the exercise. In this way the programming activity is a kind of hybrid of a web interactive where the student sees some but not all of the code. The other activities also rely on subroutines to hide uninteresting code from the user.

Since exercises 1 \& 2 are the first in the series, in order to accommodate absolute beginner programmers (who may not have completed the classical mechanics exercises), the only programming tasks in the first two exercises involves changing the charge and mass of the particle in order to determine the effect on the final speed, which they can, in the first exercise, compare to an analytic calculation from $v_{xf}^2 = v_{xi}^2 + 2 a_x \Delta x$. In the second exercise, the student changes the spacing of the plates while changing the electric field strength in order to keep the product constant (which keeps the potential difference constant, $\Delta V = - E d$) so when the slow-moving particle is accelerated by the plates it should reach the same final speed. Some students may see differences in the third digit that arise from numerical errors, which is a useful opportunity to explain that the computer is breaking up the trajectory into a number of finite time steps. Students also need to change the charge and mass of the particle again to see what the final speed of the particle is with the new plate separation and electric field. Students changed the charge and mass in the of the particle in the first exercise and it is important to reinforce their qualitative understanding that increasing the charge has the effect of increasing the final speed, and increasing mass decreases the final speed.

\section{Repulsion between two point charges}

The ``particle repulsion" exercise gives the student a code that is very similar to the particle accelerator exercise. The position and velocity update is identical. The student needs to specify the electric field that a positively charged particle experiences as it moves closer and closer to a stationary positively charged particle. This primarily involves realizing that E = k*q/(r*r) is the correct way to implement $E = k q/r^2$. Once this is done, the program determines the change in velocity during the time step, $\Delta v_x = (q E /m) \Delta t$. For simplicity, this happens in 1D and the ``target" particle is held stationary. As a result, the incoming particle slows to a stop and then turns around and accelerates in the direction it came. Students use energy conservation to calculate the distance of closest approach and this result agrees well with the program which solves the problem iteratively, rather than explicitly using energy conservation.

\begin{figure}
\includegraphics[width=3in]{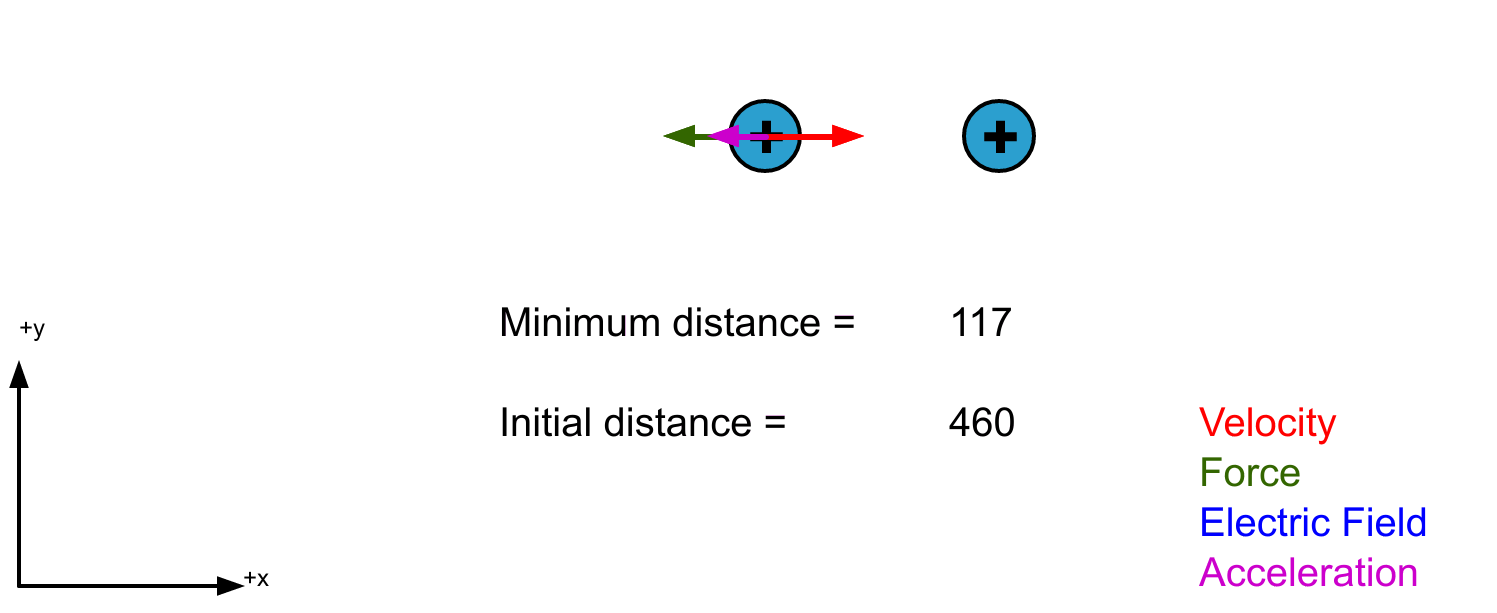}
\caption{A screenshot of the particle repulsion programming activity shortly before the incident particle comes to a stop and begins heading in the $-x$ direction.}
\end{figure}

\section{RC circuit}

\begin{figure}
\includegraphics[width=3in]{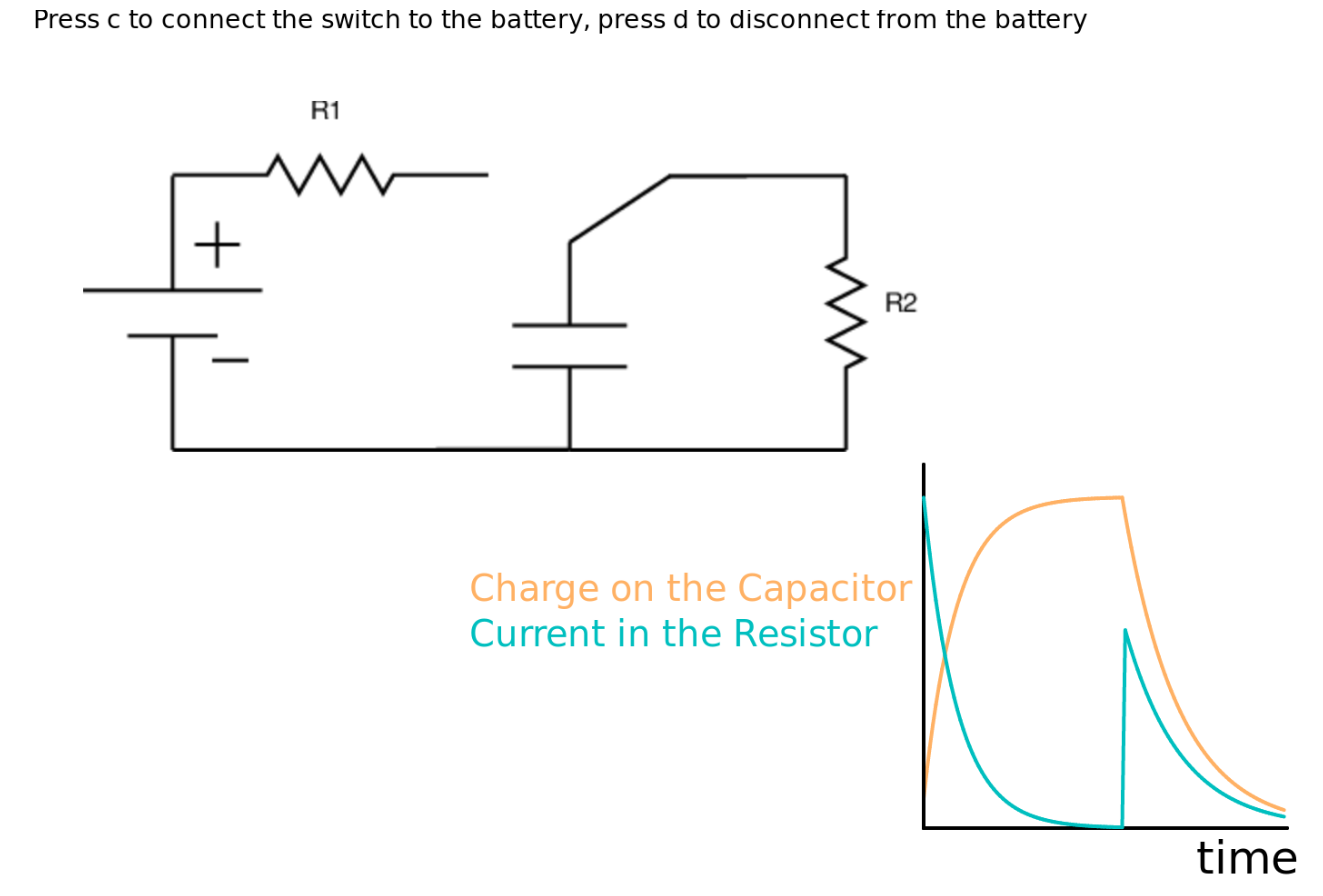}
\vspace{-0.2in}
\caption{A screenshot of the RC circuit programming activity after connecting and discharging the capacitor.}\label{fig:circuit}
\end{figure}

By necessity, the code in the RC circuit exercise is different from the previous three exercises, but the student will find that the charge and current from a discharging capacitor is updated in much the same way as the position and velocity of the particle \cite{Cromer1981}. Students are given a circuit with only one resistor ($R_2$ in Fig.~\ref{fig:circuit}), and they need to modify the code to incorporate a resistor between the capacitor and the battery ($R_1$ in Fig.~\ref{fig:circuit}). To simplify the task, the student only needs to specify the current through $R_1$ when the switch is to the left. This is $(V_{\rm batt} - V_{\rm cap})/R_1$. If the student implements this they will see the capacitor charge up and discharge realistically. Importantly, this phenomena is illustrated with an iterative computer program that uses some algebraic equations describing the circuit, instead of using calculus results or some "black box" web interactive that does not reveal any of the code. The student also changes the resistance and capacitance to see the effect on the discharge time.

\begin{figure}
\includegraphics[width=3in]{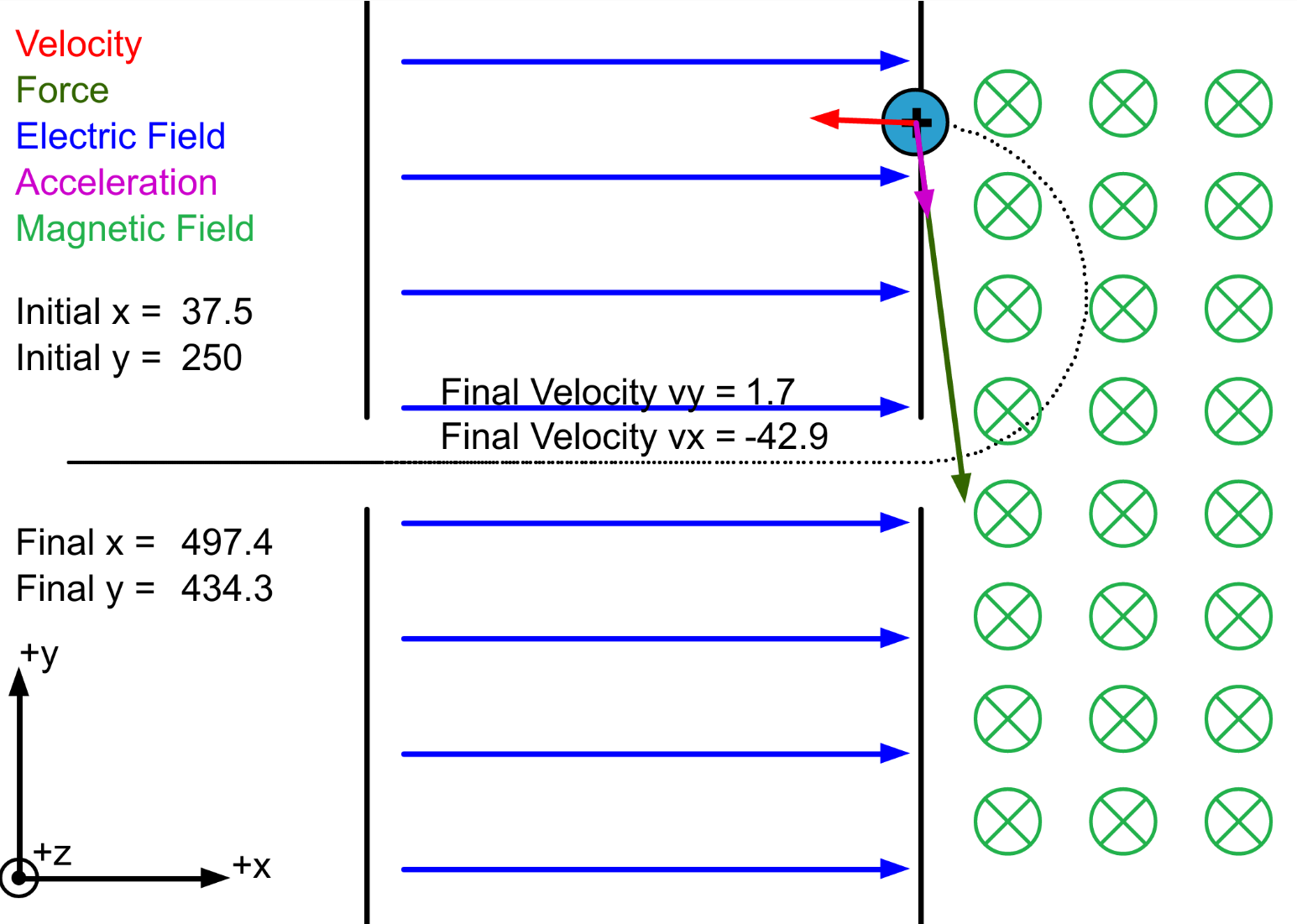}
\caption{A screenshot at the end of the completed Magnetic Force programming activity.} \label{fig:bfield}
\end{figure}

\section{Magnetic Force}

The magnetic force (a.k.a. mass spectrometer) activity is very similar to the particle accelerator activities except that there is a magnetic field into the page to the right of the plates (Fig.~\ref{fig:bfield}). The task of the student is to take the magnetic force ($F_x = q v_y B$, and $F_y = -q v_x B$) and use this to determine the acceleration and change in velocity when the particle is in the magnetic field (which is done by dividing by the mass and multiplying by the time step). When this is done, the code naturally produces a circular trajectory (Fig.~\ref{fig:bfield}), which is remarkable because a mathematical proof that particles in magnetic fields follow circular trajectories would otherwise require solving two coupled differential equations. After completing the program, the student then needs to measure the gyroradius from the simulation and compare this to the analytic formula, which is an important exercise in verifying that the code is working properly. The two should agree well, although not perfectly due to numerical errors. The code uses the same Euler-Cromer \cite{Cromer1981} method as the particle accelerator exercises, which is not a method that a research physicist would choose to solve this problem \cite{BirdsallLangdon2004}, but to show a half-circle trajectory it is perfectly adequate for students in an introductory physics course. 

\section{Wave Interference}

\begin{figure}
\includegraphics[width=2.7in]{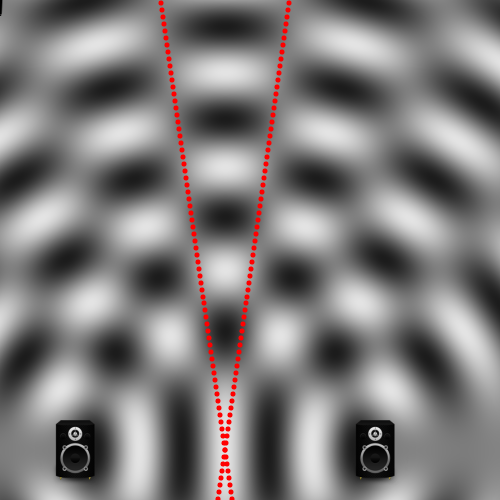}
\caption{A screenshot of the wave interference programming activity. The dotted red lines are from an analytic approximation to determine the locations of destructive interference. }\label{fig:wave}
\end{figure}

The wave interference exercise provides the student with a working interactive with two ``speakers" producing sine waves at the same frequency. The interactive is designed so that if the microphone is placed in a position of constructive interference, the interactive will produce a audible sine wave from whatever device is running it, but if instead the microphone is placed along one of the lines of destructive interference, no sound will be produced. The task of the student is to use a Talyor expansion formula ($\sqrt{1 + b/a} \approx  1 + b/(2a)$ for $a \gg b$) to simplify the path difference equation in order to conclude that the lines of destructive interference approximately follow $y = \pm [4 d / (n \lambda)]x$ for $n = 1, 3, 5...$. The student must modify the code to draw the lines for $n = 1$. This turns out to line up with the destructive interference from interactive rather well for $n = 1$ (Fig.~\ref{fig:wave}). 
Note that because the amplitude of the wave in this interactive does not decrease with distance, which greatly simplifies the analysis, this interactive is closer to ripples on a pond than speakers producing sound. The solutions for the exercise include some comments in this regard.

\section{Plans for Assessment}
\label{sec:assess}

We are developing assessments for each exercise which will be incorporated into a learning management system that will give students animated questions shortly before and after completing each exercise. The BEMA \cite{BEMA} includes a number of (static) questions relevant to particle repulsion, the deflection of particles from electric and magnetic fields and circuits.  We feel that it is important to recast these questions to have a similar look and feel as the exercises in order to avoid situations where students may be confused by the change in the style of the axes or the look of the particle, for example. Animating the assessments is likewise important, and we note that \cite{Dancy2006} found that using an animated version of the Force Concept Inventory more accurately gauged student conceptual knowledge. We expect the same will be true for electromagnetism content.


\section{Summary and Conclusion}

We present a comprehensive set of programming exercises for an introductory electromagnetism course that are appropriate for algebra-based physics at either the high school or college level where students may be absolute beginner programmers. The full set of exercises, code and solutions described here is available at \href{http://compadre.org/PICUP}{compadre.org/PICUP}.

\acknowledgements

The authors thank K. Roos and L. Englehardt from the PICUP collaboration. This project was made possible through a Connect and Collaborate Grant, a program supporting innovative and scholarly engagement programs that leverage academic excellence of OSU in mutually beneficial ways with external partners.


\bibliographystyle{apsrev}
\bibliography{main}
\end{document}